**Virtual European Solar & Planetary Access (VESPA): a Planetary Science Virtual Observatory cornerstone.**

S. Erard[1], B. Cecconi[1], P. Le Sidaner[2], C. Chauvin[2], A. P. Rossi[3], M. Minin[3], T. Capria[4], S. Ivanovski[4], B. Schmitt[5], V. Génot[6], N. André[6], C. Marmo[7], A. C. Vandaele[8], L. Trompet[8], M. Scherf[9], R. Hueso[10], A. Määttänen[11], B. Carry[12,13], N. Achilleos[14], J. Soucek[15], D. Pisa[15], K. Benson[14], P. Fernique[16], E. Millour[17]

[1]LESIA, Observatoire de Paris, Université PSL, CNRS, Sorbonne Université, Université de Paris, 5 place Jules Janssen, 92195 Meudon, France [2]DIO-VO/UMS2201 Observatoire de Paris/Université PSL/CNRS, Fr, [3]Jacobs University, Bremen, Ge [4]INAF/IAPS, Rome, It [5]Université Grenoble Alpes, CNRS, IPAG, Fr [6]IRAP/UPS/CNRS, Toulouse, Fr [7]GEOPS/CNRS/U. Paris-Sud, Fr [8]BIRA-IASB, Brussels, Be [9]OeAW, Graz, Aut [10]UPV/EHU, Bilbao, Sp [11]LATMOS/IPSL, Sorbonne Université, UVSQ, U. Paris-Saclay, CNRS, Paris, Fr [12]Université Côte d'Azur, Observatoire de la Côte d'Azur, CNRS, Laboratoire Lagrange, Fr [13]IMCCE, Observatoire de Paris, Université PSL, CNRS, Sorbonne Université, Univ. Lille, Fr [14]University College London, UK [15]Inst. of Atmospheric Physics/CAS, Prague, Cz [16]Observatoire de Strasbourg/UMR 7550, Fr, [17]LMD/IPSL, CNRS, Sorbonne Université, École Normale supérieure, Université PSL, École polytechnique, Paris, Fr

**Abstract**

The Europlanet-2020 programme, which ended on Aug 31$^{st}$, 2019, included an activity called VESPA (Virtual European Solar and Planetary Access), which focused on adapting Virtual Observatory (VO) techniques to handle Planetary Science data. This paper describes some aspects of VESPA at the end of this 4-years development phase and at the onset of the newly selected Europlanet-2024 programme starting in 2020. The main objectives of VESPA are to facilitate searches both in big archives and in small databases, to enable data analysis by providing simple data access and online visualization functions, and to allow research teams to publish derived data in an interoperable environment as easily as possible. VESPA encompasses a wide scope, including surfaces, atmospheres, magnetospheres and planetary plasmas, small bodies, heliophysics, exoplanets, and spectroscopy in solid phase. This system relies in particular on standards and tools developed for the Astronomy VO (IVOA) and extends them where required to handle specificities of Solar System studies. It also aims at making the VO compatible with tools and protocols developed in different contexts, for instance GIS for planetary surfaces, or time series tools for plasma-related measurements. An essential part of the activity is to publish a significant amount of high-quality data in this system, with a focus on derived products resulting from data analysis or simulations.

**Introduction:** Modern space borne instruments often produce large and complex datasets, especially on long-lived missions. Detailed data analysis requires new ways to handle the data, in particular to locate specific observing conditions easily and efficiently. Virtual Observatory (VO) techniques have been developed in Astronomy during the past 15 years to address similar issues; they can be adapted to this context provided they are extended to take into account the specificities of Solar System studies. The VESPA (Virtual European Solar and Planetary Access) data access system focuses on applying VO techniques and tools to Planetary Science data, and supports all aspects of Solar System science (Erard et al. 2018). VESPA was developed in the framework of the EU-funded Europlanet-2020 programme, which started Sept 1$^{st}$, 2015 and will be pursued in the recently selected Europlanet-2024 programme for another 4-year period. The objectives of VESPA are to facilitate searches both in big archives and in sparse databases, to enable data analysis by providing simple data access and on-line visualization functions, and to allow research teams to publish derived data in an interoperable environment as easily as possible. This system relies on studies and developments led in Astronomy (International Virtual Observatory Alliance, IVOA), Solar Physics (HELIO), Space Physics (SPASE), and space data archives (International Planetary Data Alliance, IPDA); it is responsive to FAIR principles and focuses on science-oriented description of data. We hereby provide a summary of achievements at the end of Europlanet-2020.

**Data services:** the VESPA architecture (Fig. 1) consists in a new data access protocol, a specific user interface querying the available data services, and intensive usage of standards and tools developed for the Astronomy VO (Erard et al. 2018 and references therein). The Europlanet data access protocol, EPN-TAP, is based on the general Table Access Protocol (TAP) associated to a set of parameters describing the content of a data service (Erard et al. 2014). These parameters describe in particular the observational and instrumental conditions for each data element, which can thus be queried by the scientific user. Data services are required to return the metadata of matching results in VOtable format, which is supported by all standard VO tools.

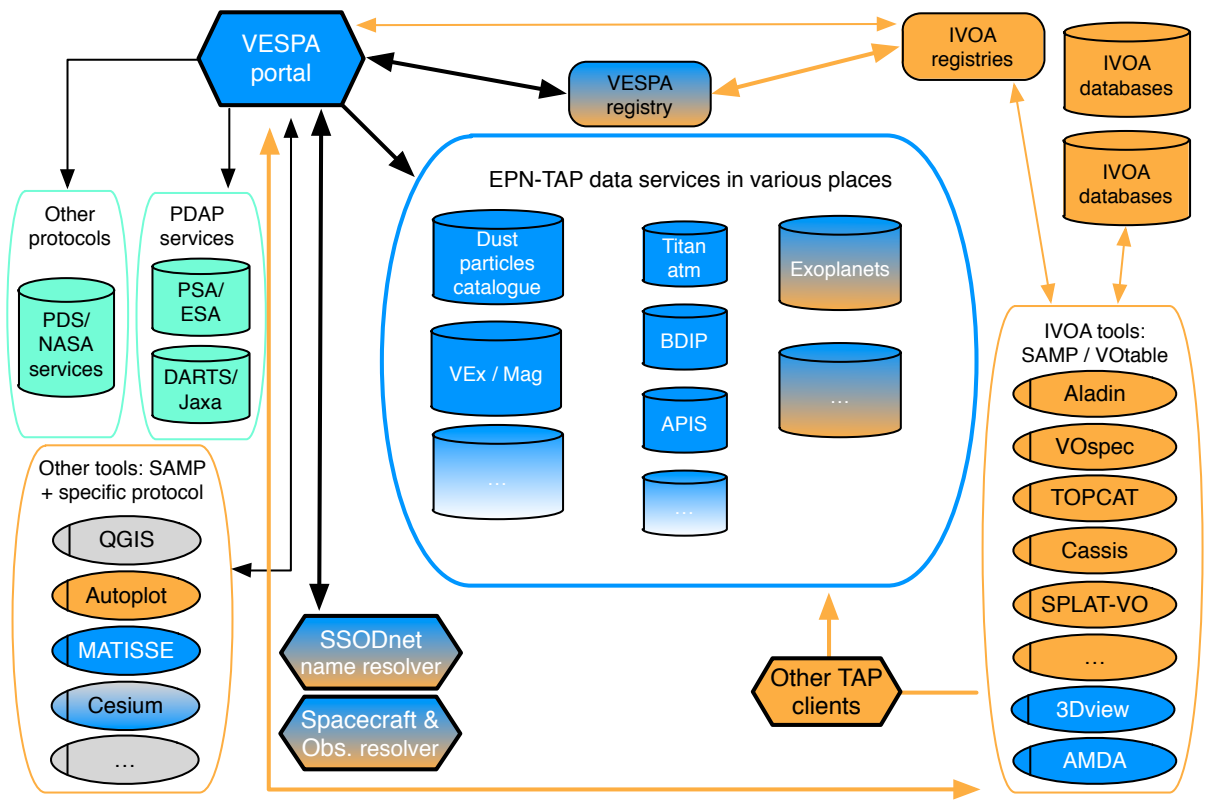

*Fig. 1: VESPA architecture and origin of developments (orange: IVOA; blue: Europlanet; cyan: PDS; grey: GIS elements)*

Data services are installed in the provider institutes and are declared in the standard IVOA registries; they are thus always visible and searchable by query interfaces. A standard procedure has been identified for design and publication, based on the DaCHS server (Demleitner et al 2014), although other solutions are possible. At the time of writing, 54 data services are publicly open, and about 15 more are being finalized[1]. They encompass a wide scope, including surfaces, atmospheres, magnetospheres and planetary plasmas, small bodies, heliophysics, exoplanets, and experimental data such as spectroscopy in solid phase. VESPA mostly focuses on derived data, typically associated to publications. Some of these derive from a significant preparation phase, e.g.: DynAstVO which computes orbital parameters for Near Earth Asteroids; APIS which provides data on planetary aurorae derived from several archives; SSHADE which currently gathers 17 experimental databases of spectroscopy of minerals, organics, ices and cosmo-materials in a consistent and very detailed format; solar feature catalogues from the HELIO EU programme; a refined and corrected version of the Robbins & Hynek Mars craters database (being published) will also be accessible, together with the original database.

Some large existing data archives also received an additional EPN-TAP interface during the programme, in particular: ESA's Planetary Science Archive (PSA), which gathers data from all European space missions (Besse et al 2018); planets and satellites data collected from the Hubble Space Telescope archive in Canada; planetary plasma data at CDPP, Toulouse, which includes many unique derived products; the IAU Minor Planet Center which distributes the properties of most minor bodies; the Encyclopaedia of Extra Solar Planets, which is the world-wide reference in the field; several ground-based solar archives (currently BASS2000 and CLIMSO) also distribute their data through VESPA.

To favour the emergence of this kind of material, VESPA has organized a yearly call to the community to select projects of interest, typically associated to a publication; 4 or 5 selected teams were invited to a one-week workshop to design and install the service in their institute[2]. Several services providing amateur astronomy data were also selected at the onset of the programme for implementation in research institutes, including PVOL (planetary images) in EHU/Bilbao (Hueso et al 2018) and RadioJove (Jupiter radio measurements) at Paris Observatory. In addition, a special type of services will gather tables of VOevents produced by alert systems in various fields (Cecconi et al. 2018b).

According to the contributory nature of the VO, any team can publish EPN-TAP services in the IVOA registry. A validator is available to certify TAP compliance, but manual checks are also performed to ensure consistency of EPN-TAP services prior to publication. By default the VESPA portal uses a local registry listing only services certified by the core VESPA team. The status of published services is monitored automatically through Nagios to inform data providers in case of problems.

**Data access:** EPN-TAP data services are best queried from the VESPA portal (Fig. 2), an optimized search interface providing specific user support (http://vespa.obspm.fr). Alternate access modes are discussed below.

In the frame of TAP, data services consist in a list of "granules", or data elements, described by a series of parameters. EPN-TAP defines a set of mandatory parameters that introduce metadata for individual granules; this scheme is inspired by the IVOA ObsTAP protocol for observational Astronomy datasets. EPN-TAP parameters provide the observational and instrumental conditions of acquisition, accounting for the specific diversity and complexity of Planetary Science: ranges along several axes (spatial, temporal, spectral, photometric), measurement type, origin of data, and several references. Localisation can be provided in various coordinate systems (celestial or planetary coordinates, either body-fixed or rotating). Time is provided in Julian days, but also as local time or season if relevant. The benefit of EPN-TAP is that all data services are described uniformly with parameters that are relevant for the science user. The VESPA portal queries all registered data services at once based on the mandatory EPN-TAP parameters, and returns individual granules matching the query; this allows in particular the user to discover unknown data content in the field of interest. In addition, thematic extensions are defined to describe new fields uniformly (e.g., lab spectroscopy). Specific parameters may also be made-up to describe individual services with more details; when querying a single service, they can be used to identify granules more precisely. Independently from VESPA and the data distribution context, the EPN-TAP Data Model (Erard et al, submitted to IVOA) can also be used or adapted to describe private databases in a consistent way, e.g. to share proprietary data inside an experiment team.

---

[1] A roadmap for service publication is maintained here:
https://voparis-wiki.obspm.fr/display/VES/EPN-TAP+Services
[2] Implementation tutorials are available: https://voparis-wiki.obspm.fr/display/VES/Implementing+a+VESPA+service

*Fig. 2: The VESPA main portal: http://vespa.obspm.fr*

The VESPA portal also supports other query systems. User queries are converted and forwarded to space agencies' PDS (Planetary Data System) archives: ESA and JAXA are queried via the PDAP protocol from IPDA, and NASA through its PDS keyword-search interface. Owing to limitations in these protocols, such queries are performed only at dataset level.

EPN-TAP services can be accessed via other search interfaces. An EPN-TAP library was developed and included in several tools (3DView, CASSIS, and AMDA) to issue direct queries from these environments. Since EPN-TAP relies on the more general TAP mechanism, EPN-TAP data services can also be accessed individually via standard TAP clients; these include general query interfaces (e.g., TAPHandle) as well as standard VO tools (e.g., TOPCAT, Aladin, etc). Programmatic access is possible using existing libraries in python (pyvo), IDL (SSW), or shell scripts (TAPsh) - making such data services handy for pipeline processing. Finally, a map-ping app prototype was also developed to explore new types of access from mobile and computer uses.

EPN-TAP tables either provide links to data files, or include the data itself when they consist in a small set of scalar quantities. A variety of tools are available to handle the data in the VO. Adequate VO tools are identified through data description parameters, which not only indicate a file format but also specify dimensions, units, and physical quantities, based on IVOA Data Models adapted for VESPA. For instance, images and spectra will open in different tools, and the spectral tools will recognize spectra in radiance or in reflectance, and handle them differently.

**Tools:** Selected metadata can be transferred from the VESPA portal to VO tools via the SAMP protocol from IVOA. Standard VO tools are connected to the VESPA portal so that they readily display metadata, e.g., spatial footprints are plotted on a 3D sphere in Aladin or Mizar; other metadata such as local time or instrument modes, can be plotted in 2D or 3D with TOPCAT (Fig. 3)[3].

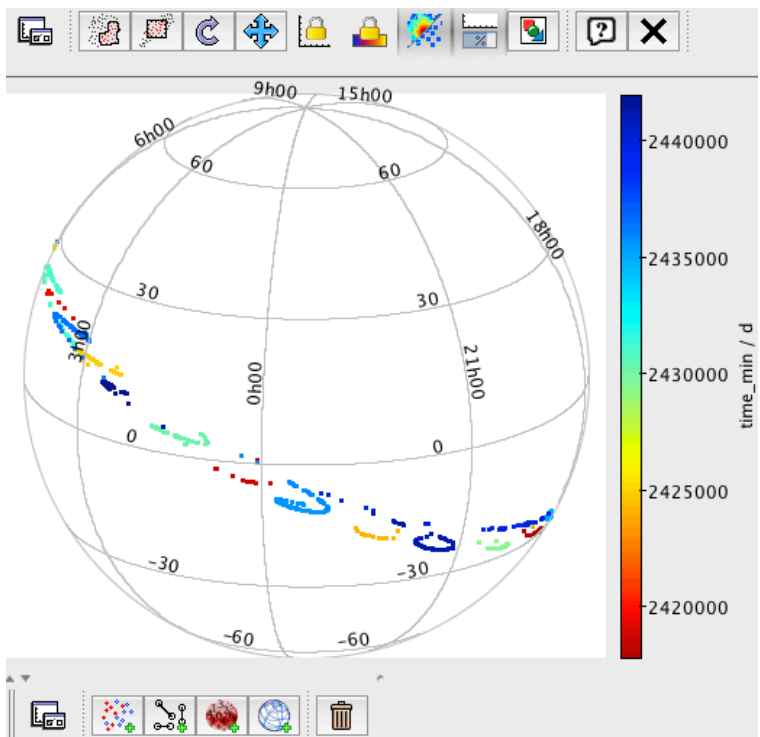

*Fig. 3: Location of Mars in the sky from historical telescopic images (1905 to 1976) displayed in TOPCAT, evidencing retrograde motion around each opposition (the colour scale provides time in Julian Days)*

The data themselves can be transferred in a similar way for display and standard analyses. Data description is used to select appropriate tools, e.g., TOPCAT handles all types of tabular data (Taylor 2017), Aladin most images and spectral cubes, CASSIS and SPLAT-VO spectra in general, 3DView displays observations of various types in 3D planetary contexts (Génot et al., 2018), Autoplot is dedicated to extracting data from long time series[4].

---

[3] see Erard et al 2018 and references therein for tools; IVOA standards are documented here: http://www.ivoa.net/documents/

Most of these tools have been updated in collaboration with their developers to support Planetary Science and specificities of Solar System data, e.g., coordinate systems on surfaces and in magnetospheres, or measurements in reflected light (Fig. 4). In some instances new tools have been developed for VESPA, such as Planetary Cesium Viewer to visualise catalogues of features on planetary surfaces; APERICubes to slice PDS3 spectral cubes (Savalle et al. 2016); iPEC-MAN to analyse multi-dimensional measurements of planetary electromagnetic fields (Pisa et al 2017). Some existing non-VO tools have been provided with a SAMP interface to exchange data in a VO context and can be integrated in workflows, e.g.: ImageJ which now provides conversions for many formats, as well as image processing functions to the VO; QGIS, an Open Source GIS application; Autoplot (see below); MATISSE to plot data on 3D shape models (Longobardo et al., 2018). Finally, specific web tools developed as part of larger data services have been made accessible for use with external data, e.g. AMDA for time series at CDPP (Génot et al. 2014), or the new SSHADE service for lab spectroscopy (Schmitt et al. 2018).

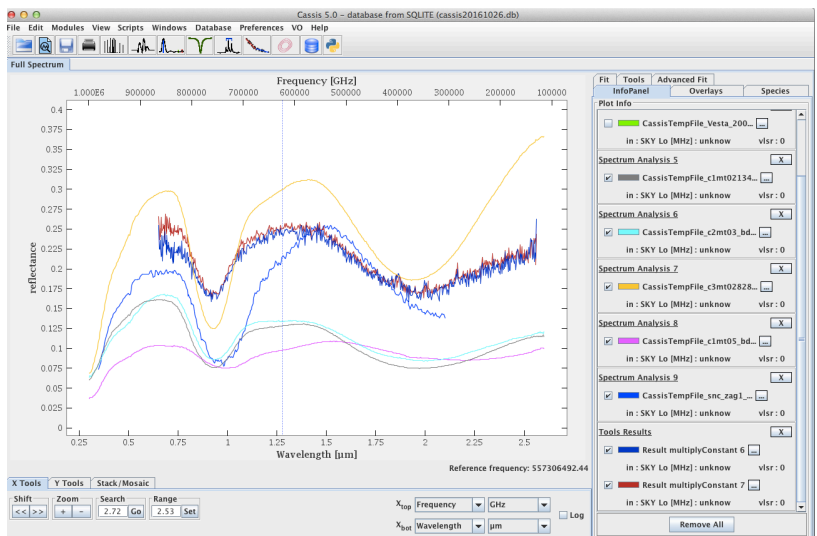

*Fig. 4: NIR telescopic spectra of 4 Vesta (from the m4ast service) compared to basaltic meteorites from the PDS spectral library in CASSIS, on a reflectance scale.*

TOPCAT can easily integrate sparse surface observations (e.g. from a point spectrometer) using the healpix tesselation system, while Aladin can produce multi-resolution maps (HiPS) from large datasets, which allow for smooth and fast change of scale (Fernique et al., 2015). Currently, 60 planetary maps from USGS have been converted to HiPS and are available from the Aladin data tree. The same technique applied to large panoramas from planetary landers provides a very exciting way to navigate within such images, by smoothly changing from the global picture to the highest local details.

A significant activity is the development of a connection between the VO world and Geographic Information Systems (GIS). In a first step, EPN-TAP services were used to provide links as queries to WMS or similar services, i.e. using different, non-VO, access protocols. Traditionally, such links are only handled in GIS applications such as the open source QGIS. While the intermediate VO layer allows for powerful search functions in the databases, cross-examinations with other datasets remains difficult because of the variety of query systems and image formats (e.g., Hare et al 2018). In a second step, the goal was to provide bridges between these two worlds, so that VO (e.g., FITS) and GIS (e.g., GeoTIFF) images can be displayed in all applications. This is done on one hand by providing improved georeferentiation support in FITS headers and conversion routines in the GDAL library, which is widely used to import data in applications such as QGIS (Rossi et al. 2016, Marmo et al. 2018, Fig. 5), on the other hand with new QGIS plug-ins to add SAMP connectivity (Minin et al. 2019).

---
[4] Practical tutorials are available here: http://www.europlanet-vespa.eu/tutos.shtml

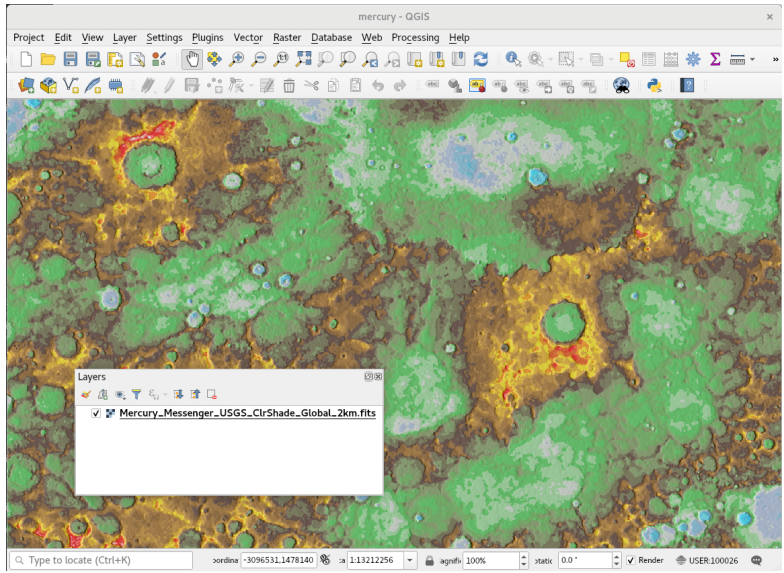

*Fig. 5: A MESSENGER multispectral image of Mercury converted to GeoFITS format is directly opened in QGIS and plotted on a coordinate grid through the updated GDAL library.*

A similar situation applies to time series depicting radio emission of the planets. A protocol of choice in this case is das2, which allows the distribution of data with adaptive temporal resolution. Data services are responsive to EPN-TAP but provide data as queries to such servers, the results of which can be fetched via SAMP to the Autoplot tool for display (Cecconi et al. 2018a, Fig. 6).

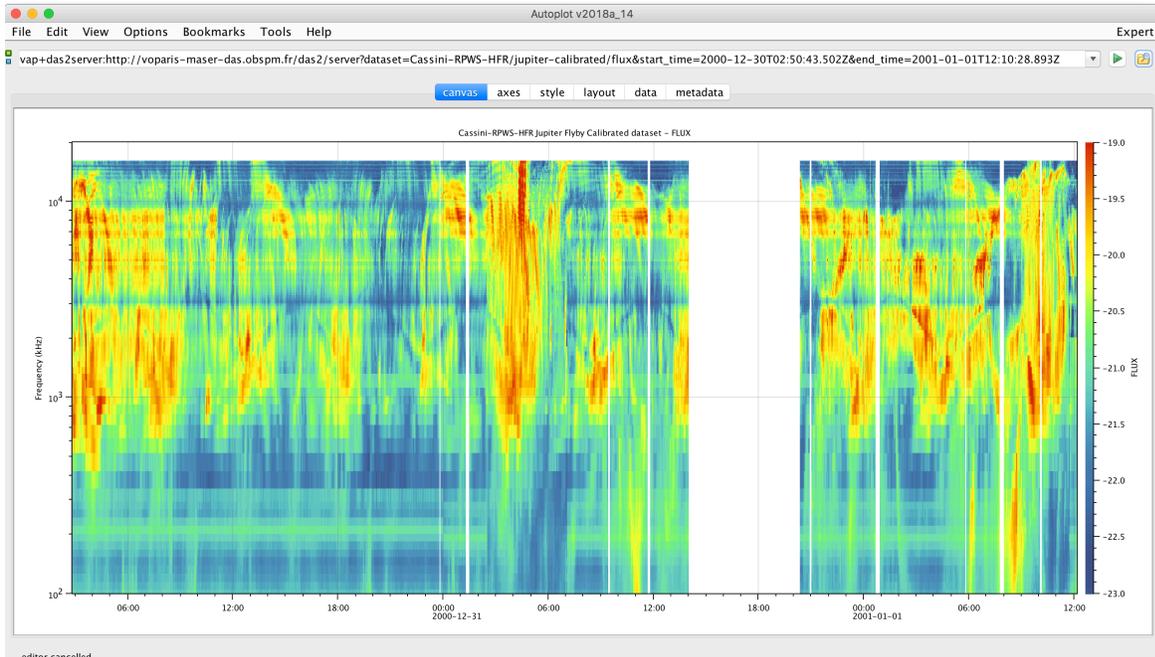

*Fig. 6: Cassini RPWS/HFR Jupiter Flyby Calibrated Dataset displayed in Autoplot, using the das2server interface.*

As far as 2D data are concerned, VESPA makes use of two IVOA protocols to handle footprints. The first one is the STC-S standard (used in particular by the ObsTAP protocol) which provides oriented contours; the second one is the Multi-Order Coverage (MOC, healpix based) used e.g. in Aladin, TOPCAT, and Mizar. Both standards can be used to issue powerful searches on intersections or inclusions, and to select objects within arbitrary footprints (Fig. 7).

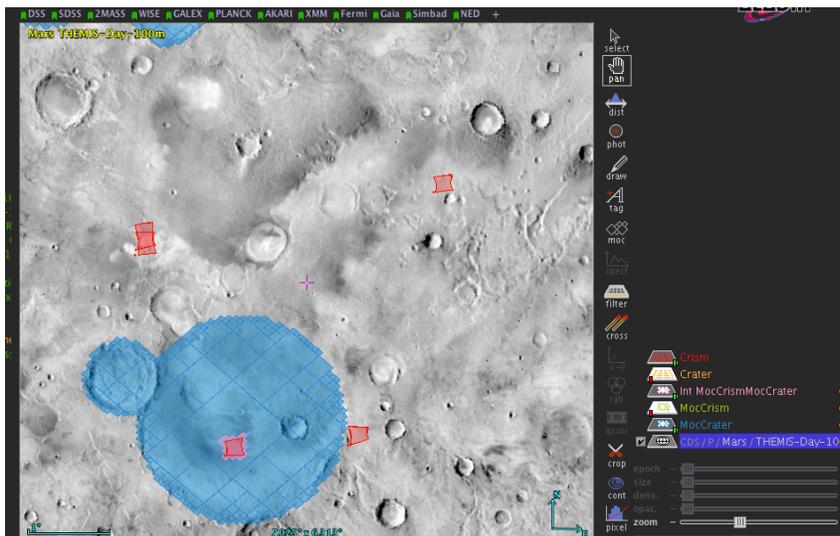

*Fig. 7: Selection of Mars Reconnaissance Orbiter CRISM spectral cubes located in large Martian craters, based on MOC in Aladin (over THEMIS daytime HiPS).*

**Simulation services:** another important goal for VESPA is to connect online computation services with an interface similar to that of data services, so as to compare observations and simulations more routinely. This activity has obvious applications, e.g., for radiative transfer in planetary atmospheres or for magnetospheres, but also to connect ephemeris systems (e.g. Miriade) with data services. The datalink protocol of IVOA is used to call web services with parameters retrieved from existing data services, e.g.: Mars-Express SPICAM vertical profiles are linked to simulations from the Mars Climate Database; HST data to physical ephemerides from Miriade. More direct ways to launch simulations on demand are being set up (Trompet et al. 2017, Cecconi et al. 2018a). Independently, a new function called ViSiON has been developed on Miriade to help plan observations of planetary objects from arbitrary locations (Carry & Berthier 2018).

An additional aspect is to provide online low-level computation functions, e.g. averaging, resampling, deconvolution of actual data. This is currently supported only to some extent by standard VO tools and ImageJ; in addition, higher level processing such as retrieval of Hapke parameters from surface spectra, multivariate analyses, etc, would also be beneficial and are being investigated.

**Building a community:** Hands-on sessions have been organized twice a year at EGU and EPSC conferences in Europe to support new users, as well as contributions to similar workshops in Astronomy. Besides, a procedure has been developed to set up data services easily and with limited resources: a complete service can be installed within a week with no prior experience; the most challenging part is to provide adequate data description, which of course may be demanding for complex archives. This is expected to foster the installation of data services in research institutes, in particular to distribute derived data related to a publication. In parallel, discussions are held with big data providers, in particular space agencies in the frame of the IPDA. Finally, a Solar System Interest Group has been initiated in the IVOA in 2017, to which several VESPA partners contribute.

**Prospects:** At the end of Europlanet-2020, VESPA provides a consistent data distribution infrastructure open to contributions from the community. The data description system makes it possible to identify specific instrumental or observational conditions in datasets encompassing many areas of Solar System studies. The search system connects with powerful display and analysing tools. Collaborations started in several areas (in particular exoplanets, heliophysics, experimental spectroscopy, surface studies, small bodies, and radio observations of magnetized planets) will help coordinate these fields and make future data services interoperable.

During the coming Europlanet-2024 programme, VESPA will continue to set up new data services and to connect existing ones, e.g. by setting up bridges with PDS4 archives at NASA and ESA, and serving other space agencies as well. Connections between PDS4 and EPN-TAP dictionaries would make all PDS metadata searchable from the VESPA portal and vice versa. Another goal is to connect Solar System data present in astronomical VO catalogues, starting with the VizieR database (Ochsenbein et al., 2000). Finally, experimental work performed by other Europlanet activities will also be distributed in VESPA.

Using a light and distributed data system (as opposed to heavy and expensive data centres) to publish contributions from individual research teams is efficient, in particular in the context of Open Science and incitation to make research data widely available. However, there are drawbacks related to sustainability. Services may disappear or become obsolete whenever local interest vanishes, which often relies on a single person. A possible workaround, in addition to getting space and telescope agencies involved, is to rely on a series of regional centres taking care of existing services that are no longer maintained by the original publishers; in Europlanet-2024, Paris and Trieste observatories and the Heidelberg University, three major repositories of VO services, will act as regional VESPA hubs securing published resources. This is made easier if the data services are actually located on a cloud, rather than in the institutes; this may also provide a workaround to strict IT policy in some institutes. An obvious evolution will therefore be to consider the new EU-funded European Open Science Cloud (EOSC) being developed in the H2020 framework. Finally, other activities in Europlanet-2024 will develop state-of-the-art processing techniques in several fields, in particular related to planetary mapping and Machine Learning. These activities will make use of VESPA data services, and will distribute their results in VESPA. Such developments are retained necessary in view of future space missions such as BepiColombo and JUICE.

**Acknowledgements:**
The Europlanet-2020 Research Infrastructure project has received funding from the European Union's Horizon 2020 research and innovation programme under grant agreement No 654208. Support from Paris Astronomical Data Centre (PADC) is acknowledged.
VESPA web site: http://www.europlanet-vespa.eu

**References:**

Besse, S., Vallat, C., Barthelemy, M., Coia, D., Costa, M., De Marchi, G., Fraga, D., Grotheer, E., Heather, D., Lim, T., Martinez, S., Arviset, C., Barbarisi, I., Docasal, R., Macfarlane, A., Rios, C., Saiz, J., and Vallejo, F. 2018 ESA's Planetary Science Archive: Preserve and present reliable scientific data sets. *Planet. Space Sci.* 150, 131–140, 10.1016/j.pss.2017.07.013

Carry, B. and Berthier, J. 2018 ViSiON: Visibility Service for Observing Nights. *Planet. Space Sci.* 164, 79–84. 10.1016/j.pss.2018.06.012

Cecconi B., Le Sidaner P., Savalle R., Bonnin X., Zarka P., Louis C., Coffre A., Aicardi S., Lamy L., Denis L., Grießmeier J.-M., Faden J., Piker C., André N., Génot V., Erard S., Mafi J., King T., Sharlow M., Sky J., Demleitner M. 2018a MASER: A Toolbox for Low Frequency Radio Astronomy. EPSC, Berlin 17-21 Sept 2018. id.EPSC2018-822

Cecconi, B., Le Sidaner P., André N., Tomasik L., Gangloff M., Marmo C. 2018b VOevent for Sun-Earth and planetary space weather. 42nd scientific assembly of the COSPAR, Pasadena, USA. Abst. PSW.4-0008-18.

Demleitner, M., Neves, M. C., Rothmaier, F., and Wambsganss, J. 2014 Virtual Observatory publishing with DaCHS. *Astron. and Comput.* 7, 27–36. 10.1016/j.ascom.2014.08.003

Erard S., Cecconi B., Le Sidaner P., Rossi A.P., Capria M.T., Schmitt B., Génot V., André N., Vandaele A.C., Scherf M., Hueso R., Määttänen A., Thuillot W., Carry B., Achilleos N., Marmo C., Santolik O., Benson K., Fernique P., Beigbeder L., Millour E., Rousseau B., Andrieu F., Chauvin C., Minin M., Ivanoski S., Longobardo A., Bollard P., Albert D., Gangloff M., Jourdane N., Bouchemit M., Glorian J., Trompet L., Al-Ubaidi T., Juaristi J., Desmars J., Guio P., Delaa O., Lagain A., Soucek J., and Pisa D. 2018 VESPA: A community-driven Virtual Observatory in Planetary Science. *Planet. Space Sc.* 150, 65-85, 10.1016/j.pss.2017.05.013. ArXiv 1705.09727

Erard S., Cecconi B., Le Sidaner P., Berthier J., Henry F., Molinaro M., Giardino M., Bourrel N., André N., Gangloff M., Jacquey C., and Topf F. 2014 The EPN-TAP protocol for the Planet-

ary Science Virtual Observatory *Astron. & Comput.* 7-8, 52-61, 10.1016/j.ascom.2014.07.008. ArXiv 1407.5738

Erard S., Cecconi B., Le Sidaner P., Deimletner M. 2019 EPN-TAP: Publishing Solar System Data to the Virtual Observatory. IVOA Working Draft

Fernique, P., Allen, M. G., Boch, T., Oberto, A., Pineau, F.-X., Durand, D., Bot, C., Cambrésy, L., Derriere, S., Genova, F., and Bonnarel, F. 2015 Hierarchical progressive surveys. Multi-resolution HEALPix data structures for astronomical images, catalogues, and 3-dimensional data cubes. *Astron. & Astrophys.* 578 A114. 10.1051/0004-6361/201526075

Génot, V., André, N., Cecconi, B., Bouchemit, M., Budnik, E., Bourrel, N., Gangloff, M., Dufourg, N., Hess, S., Modolo, R., Renard, B., Lormant, N., Beigbeder, L., Popescu, D., and Toniutti, J.-P. 2014 Joining the yellow hub: Uses of the Simple Application Messaging Protocol in Space Physics analysis tools. *Astron. and Comput.* 7, 62–70. 10.1016/j.ascom.2014.07.007.

Génot, V., Beigbeder, L., Popescu, D., Dufourg, N., Gangloff, M., Bouchemit, M., Caussarieu, S., Toniutti, J.-P., Durand, J., Modolo, R., André, N., Cecconi, B., Jacquey, C., Pitout, F., Rouillard, A., Pinto, R., Erard, S., Jourdane, N., Leclercq, L., Hess, S., Khodachenko, M., Al-Ubaidi, T., Scherf, M., Budnik, E. 2018 Science data visualization in planetary and heliospheric contexts with 3DView, *Planet. Space Sc.*, 150, 111-130, 10.1016/j.pss.2017.07.007

Hare, T. M., Rossi, A. P., Frigeri, A., and Marmo, C. 2018 Interoperability in planetary research for geospatial data analysis. *Planet. Space Sci.* 150, 36–42. 10, 1016/j.pss.2017.04.004. arXiv 1706.02683.

Hueso R., Juaristi J., Legarreta J., Sánchez-Lavega A., Rojas J. F., Erard S., Cecconi B. 2018 The Planetary Virtual Observatory and Laboratory (PVOL) and its integration into the Virtual European Solar and Planetary Access (VESPA). *Planet. Space Sc* 150, 22-35. 10.1016/j.pss.2017.03.014. ArXiv 1701.01977

Longobardo, A., Zinzi, A., Capria, M. T., Erard, S., Giardino, M., Ivanovski, S., Fonte, S., Palomba, E., Di Persio, G., and Antonelli, L. A. 2018 Production and 3D visualization of high-level data of minor bodies: The MATISSE tool in the framework of VESPA-Europlanet 2020 activity. *Adv. Space Res.* 62, 2317–2325. 10.1016/j.asr.2017.12.001.

Marmo C., Hare T.M., Erard S., Minin M., Pineau F.-X., Zinzi A., Cecconi B., Rossi A. P. 2018 FITS format for planetary surfaces: definitions, applications and best practices. *Earth and Space Science* 5, 640-651 (special section *Planetary Mapping: Methods, Tools for Scientific Analysis and Exploration*). 10.1029/2018EA000388

Minin M., A. P. Rossi, R. Marco Figuera, V. Unnithan, C. Marmo, S. Walter, M. Demleitner, P. Le Sidaner, B. Cecconi, S. Erard, T. M. Hare (2019) Bridging the gap between Geographical Information Systems and Planetary Virtual Observatory. *Earth and Space Science* 6, 515– 526 (special section *Planetary Mapping: Methods, Tools for Scientific Analysis and Exploration*). 10.1029/2018EA000405

Ochsenbein, F., Bauer, P., and Marcout, J. 2000 The VizieR database of astronomical catalogues. *Astron. and Astrophy. Suppl.* 143, 23–32. 10.1051/aas:2000169

Píša D., Santolík O., Souček J., and Taubenschuss U. 2017 Implementation of the interface for sPECtral Matrix ANalyzer (iPECMAN). European Planetary Science Congress, Riga, 17-21 Sept 2017. EPSC2017-540.

Rossi A. P., Hare T., Baumann P., Misev D., Marmo C., Erard S., Cecconi B., and Marco Figuera R. 2016 Planetary Coordinate Reference Systems for OGC Web Services. *Lunar and Planetary Science Conference* 47 1422.

Savalle R., Erard S., and Le Sidaner P. 2016 APERICubes: an on-line astronomical and plan-

etary ergonomic research interface for spectral cubes. *ADASS* XXVI, abstract 31316, Trieste, Italy 16 - 20 October 2016.

Schmitt B., Bollard P., Garenne A., Albert D., Bonal L., Poch O. and the SSHADE consortium partners 2018 SSHADE: the European solid spectroscopy database infrastructure. European Planetary Science Congress 2018, Berlin, id.EPSC2018-529.

Taylor, M. 2017 TOPCAT: Working with Data and Working with Users. arXiv 1711.01885.

Trompet L., Vandaele A.C., Geunes Y., Mahieux A., Wilquet V., Chamberlain S., Robert S., Thomas I., Erard S., Cecconi B., and Le Sidaner P. 2017 IASB-BIRA contribution to VESPA for planetary aeronomy studies. European Planetary Science Congress, Riga, 17-21 Sept 2017. EPSC2017-181.